\documentclass[12pt]{iopart}
\usepackage{graphicx} 
\usepackage{stfloats}

\begin{document}
\title[]{Intermittent excitation of parametric instabilities enhanced by synergistic effect in broadband laser plasma interaction}
\author{Q K Liu$^{1,2}$, E H Zhang$^{1,2}$\footnote{Q K Liu and E H Zhang contributed equally to this work.},
W S Zhang$^1$, H B Cai$^1$, Y Q Gao$^3$ and S P Zhu$^{1,2}$}

\address{$^1$ Institute of Applied Physics and Computational Mathematics, Beijing 100094, China}
\address{$^2$ Graduate School of China Academy of Engineering Physics, Beijing, 100088, China}
\address{$^3$ Shanghai Institute of Laser Plasma, Shanghai, 201800, China}
\eads{\mailto{cai\_hongbo@iapcm.ac.cn}, \mailto{zhu\_shaoping@iapcm.ac.cn}}
\begin{abstract}
   A new evolution pattern for parametric instabilities in the non-linear stage driven by a broadband laser is studied with kinetic particle-in-cell simulations. It is found that an intermittent excitation of parametric instabilities caused by the high-intensity pulses, which are generated by the laser intensity variation of the broadband laser. The synergism between these high-intensity pulses reduces the Landau damping through the co-propagation of the electron-plasma waves and hot electrons, promoting the transition from convective to absolute stimulated Raman scattering. These intermittently excited parametric instabilities, especially absolute instability, will produce over-expected hot electrons compared with linear theory. However, comparisons between the energy of scattered lights for various bandwidth lasers indicate suppression of parametric instabilities with large laser bandwidths. Our works provided an opportunity to advance the understanding of the broadband laser-plasma interaction.
    
\end{abstract}


\submitto{\NF}

\section{Introduction}
In indirect inertial confinement fusion (ICF) experiments, the laser beamlets incident the hohlraum through the entrance holes and strike the inside of the hohlraum to generate x-rays.\cite{RN0,RN1,RN2,NF1}. These lasers can motivate various parametric instabilities, which bring troubles to the ICF ignition. The high-intensity electron plasma waves (EPWs) generated by stimulated Raman scattering (SRS) and two-plasmon decay (TPD) are particularly problematic because they produce hot electrons\cite{RN3,RN4,NF2,NF3,RN5,RN6}, which potentially preheat the fuel and prevent compression of the capsule. Several techniques have been proposed to suppress SRS and TPD, such as external DC magnetic fields\cite{RN7}, beam smoothing methods\cite{RN8,RN9,RN10}, and broadband laser. \par
It has long been known that broadband laser can reduce the growth rate and increases the thresholds for these parametric instabilities. Thomson and Karush analyzed the suppression of parametric instabilities driven by a broadband laser with a discrete spectrum\cite{RN11}, and their model was extended to continuous spectrum by \emph{Pesme} \etal\cite{RN12}. Their results show that the growth rates of parametric instabilities vary inversely with the bandwidth of the broadband laser in the linear stage. Recently, numerical simulation tools have been widely used in broadband laser-plasma interaction due to the improvement of super-computer performance. \emph{Zhao} \etal numerically investigated several parametric instabilities motivated by broadband laser\cite{RN13}. They also proposed a new laser model called decoupled broadband laser, which can suppress these instabilities more effectively\cite{RN14,RN15}. \emph{Follet} \etal found that the broadband laser significantly increases the absolute threshold for SRS and TPD\cite{RN16,RN17}. \emph{Zhou} \etal focused on the dynamics of the non-linear stage of SRS\cite{RN18}. They found that the tolerance of a higher limit of frequency shift causes the enhancement of SRS in the non-linear stage. However, the researches to date have tended to use large bandwidth lasers and focus on the overall effects of the broadband laser. Considering the recent improvement of broadband laser facilities\cite{GAO1,RN20}, the research based on the performance of current laser facilities is valuable, which means constructing a broadband laser model according to the characteristics of these facilities and using relatively narrow bandwidth. Furthermore, it should be noted that the coherent time of broadband laser is short, and the change of coherence of different frequency components can affect its intensity. Thus, the growth process of parametric instabilities is different from that of a normal laser case. So far, there has been little discussion about the evolution of parametric instabilities driven by a broadband laser model based on the current laser facilities.\par
This work presents a numerical study of the growth process for SRS and TPD driven by a broadband laser model based on KunWu\cite{GAO1}. It was found that the variation of broadband laser intensity was caused by the change of coherence of different frequency components. The variation generates random high-intensity pulses, leading to an intermittent excitation of these instabilities. The non-linear transition from convective to absolute SRS is observed due to the synergism between these pulses, which leads to a burst of instantaneous reflectivity and produces a large number of hot electrons. The effect of bandwidth on the evolution of SRS was also explored. We found that the average duration of pulses is inversely proportional to the laser bandwidth. With a large bandwidth, the duration of pulses falls below the characteristic time of SRS, thus the growth process of SRS is frequently interrupted. To the best of our knowledge, this is the first time that the collective behavior in parametric instabilities driven by the broadband laser has been revealed. The physical mechanism studied in this work may be helpful to explain the experimental results on current and future broadband laser facilities.

\section{Theoretical Analysis}
\subsection{Numerical Model of Broadband Laser}
It is proposed that there are several schemes to model a broadband laser, such as the Kubo-Anderson process (KAP) and a mosaic of beamlets\cite{RN19}. Here we consider a numerical model where it is composed of many coherent beamlets, each with a different frequency and an additional phase:
\begin{equation}
   E_{sum}=\frac{1}{\sqrt{N}}\sum^{N}_{i=1}E_i\cos{(\omega_i t+\phi_i)} \label{1},
\end{equation}
where $E_i$ is the electric field amplitude of the beamlet $i$. $N$ is the number of beamlets, which typically around a few hundreds, and $\phi_i$ is a random phase in $[-\pi/2,\pi/2]$. The frequency $\omega_i$ can be defined within a certain bandwidth $\Delta\omega$ around the central frequency $\omega_0$. We let $\omega_i$ be randomly selected within $[\omega_0-\Delta\omega/2,\omega_0+\Delta\omega/2]$, and we use the flat power spectrum\cite{RN16}, which means all beamlets have the same $E_i$, to match the condition of the broadband laser facility KunWu\cite{GAO1}. $1/\sqrt{N}$ is a normalization factor for the flat power spectrum.

\subsection{Variation of Broadband Laser Intensity}

\begin{figure}
    \centering
    \includegraphics[width=0.96\textwidth]{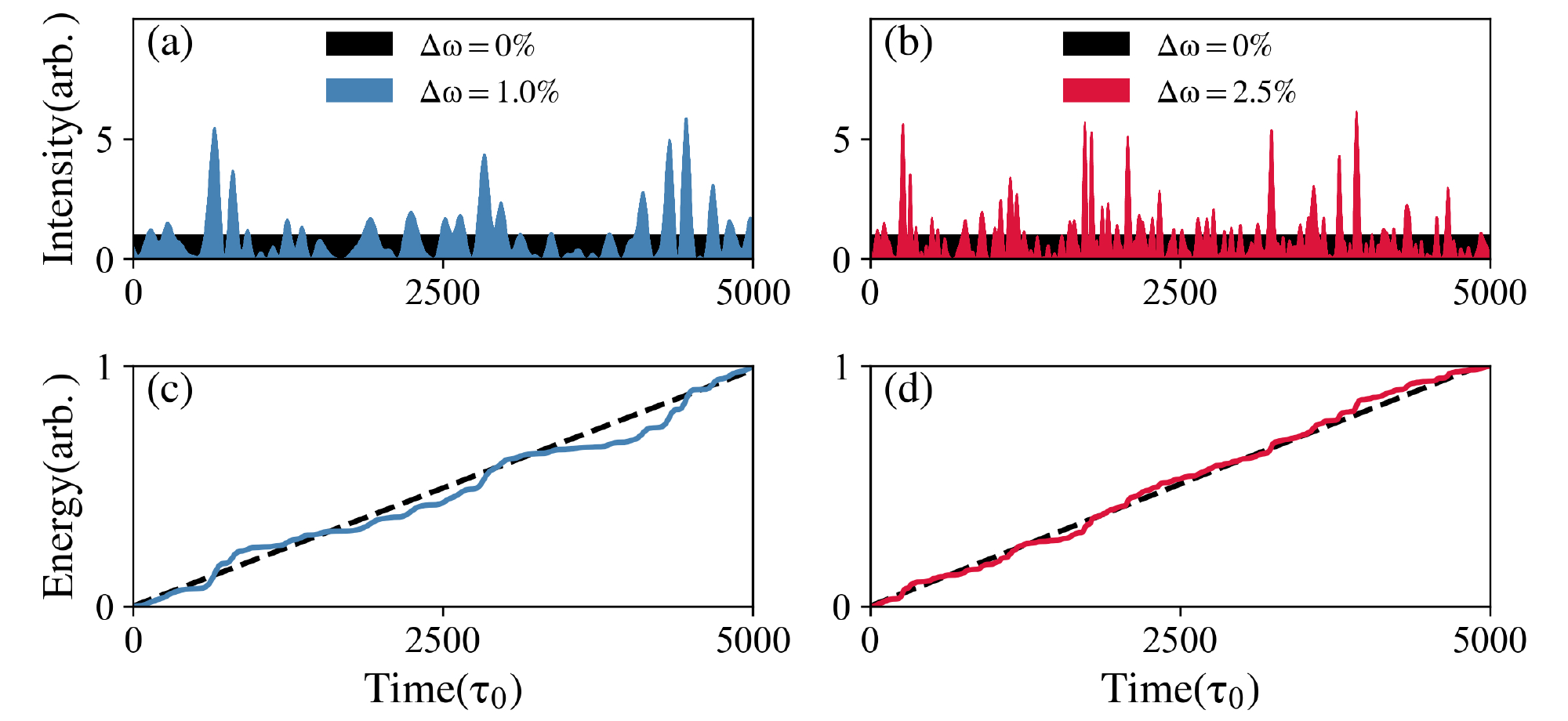}
    \caption{Temporal envelopes of the incident intensity for both normal (black area) and broadband laser (blue/red area), where the bandwidth of broadband laser is $1.0\%$ (a) and $2.5\%$ (b). The input energy for both normal (black dash line) and broadband laser (blue/red line), where the bandwidth of broadband laser is $1.0\%$ (c) and $2.5\%$ (d). }
    \label{fig:1}
 \end{figure}

The model above ensures that the input energy of the broadband laser is approximately the same as that of the normal laser, as shown in Figure 1\;(c) and (d). However, the incident intensity of the broadband laser behaves quite differently from that of the normal laser. Figure 1\;(a) and (b) show temporal envelopes of the incident intensity of two broadband lasers for bandwidths $\Delta\omega/\omega_0=1.0\%$ and $\Delta\omega/\omega_0=2.5\%$, with a normal laser as a control. The broadband laser intensity varies significantly in the form of a series of high-intensity pulses in both figures. The intensity for the strongest pulse is six times higher than the normal laser. We also noticed that the duration of these pulses varies inversely with the bandwidth. \par

The intensity variation is caused by the change of coherence of different frequency beamlets. Therefore, the average duration of the pulses can be estimated with the coherence time of the broadband laser. Using the Wiener-Khinchin theorem\cite{RN21}, which states that the correlation function and the power spectrum of a wide-sense stationary random process form a Fourier transform pair, we give the coherence time:
\begin{equation}
   \tau_{c}=2\pi/\Delta\omega \label{2}
\end{equation}
 for flat spectrum. In our simulation, a laser with $1.0\%$ bandwidth has an average pulse duration $\tau_{avg}=111fs$ and a coherent time $\tau_c=117fs$, which proves that the coherence time is a good approximation of the average duration for the pulses.

 \section{PIC Simulations and Discussions}
Previous studies have shown that laser intensity plays an important role in the evolution of parameter instabilities. Therefore, the variation of a broadband laser intensity will cause some novel phenomena. A set of numerical simulations have been carried out to investigate the effects of intensity variation by using the ASCENT code\cite{RN22}. In the following simulations, the space and time are normalized by the laser wavelength in vacuum $\lambda_0$ and the laser period $\tau_0$.

\subsection{Intermittent Excitation of SRS}

We start with the 1D case to devote to SRS. In order to investigate the SRS driven by a broadband laser, two typical simulations have been performed. One simulation uses a broadband laser with a bandwidth $\Delta\omega/\omega_0=2.5\%$, and the other uses a normal laser as a control. Both lasers are linearly polarized with the wavelength $\lambda_0=351nm$ and the intensity $I_0=2.8\times 10^{15} W/cm^2$. The simulation box contains an un-dense plasma slab with sharp edges normal to the direction of incidence of the laser. The slab width is $80\lambda_0$, and on each side of the slab, there is a $10\lambda_0$ vacuum region. The plasma has a uniform density $n_e=0.13n_c$ and an electron temperature $T_e=3 keV$. There are 50 grid points per $\lambda_0$ and 4000 particles per grid point inside the slab, which yield a total of $1.6\times 10^8$ particles. One simulation lasts $7000\tau_0$, or equivalently $8.2ps$. Previous simulations have found significant levels of backscattering with similar parameters, and it should be noticed that kinetic effects are supposed to be important for SRS growth in this parameter regime, which $k\lambda_d\approx0.32$.\par

\begin{figure}
   \centering
   \includegraphics[width=0.96\textwidth]{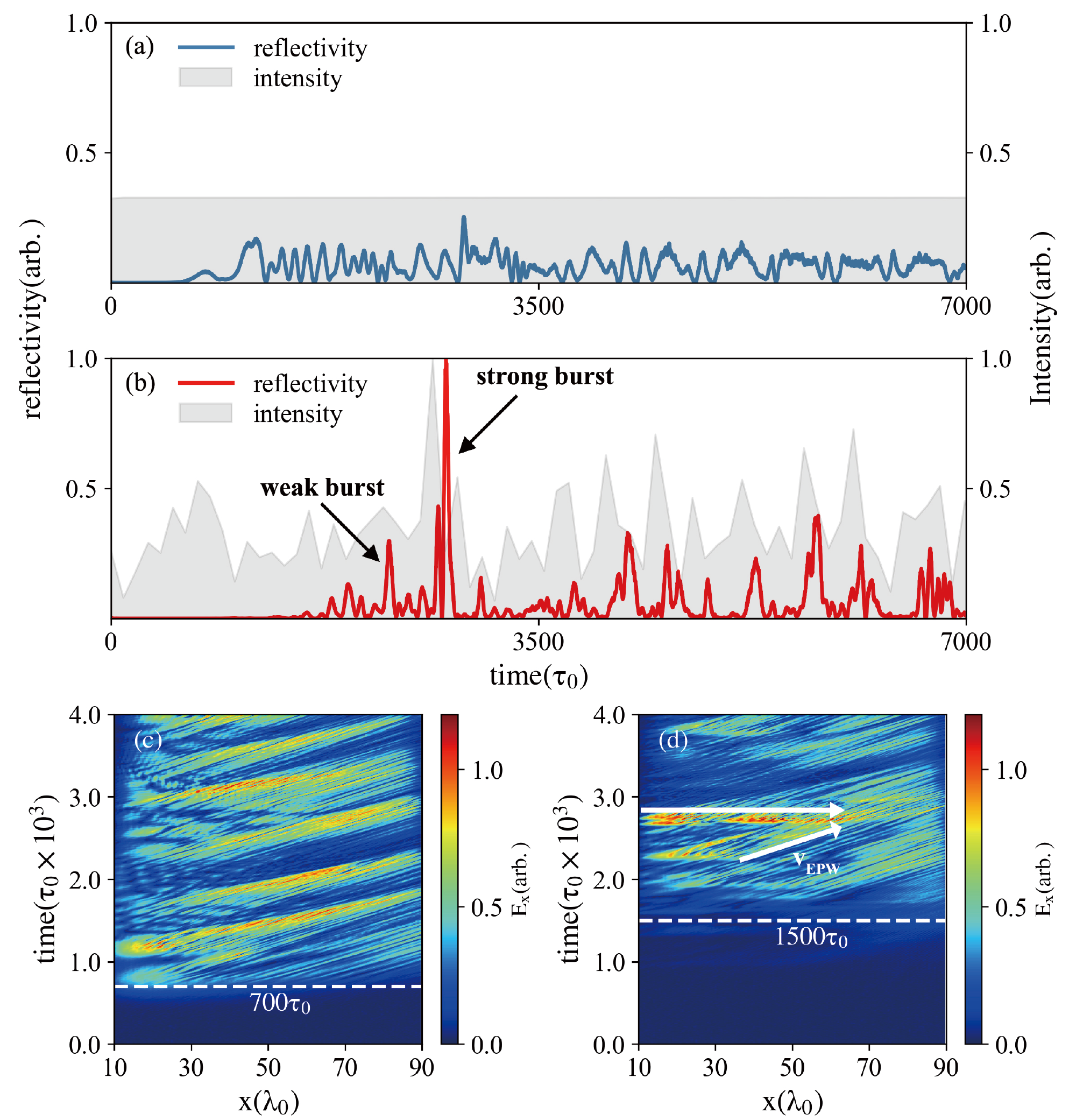}
   \caption{The average intensity ($I_{in}$) and instantaneous reflectivity observed at the left simulation boundary driven by a normal laser (a) and a broadband laser (b), both reflectivity and intensity normalized to the max reflectivity $r_{max}$ and the max intensity $I_{max}$ in the broadband laser case. Time vs. space of the electron plasma waves ($E_x$) for the normal laser case (c) and the broadband laser case (d).}
   \label{fig:2}
\end{figure}

Figure 2\;(a) and (b) show temporal evolutions of reflectivity and laser intensity for the two simulations. The intensity is averaged using a window in the time of $100\tau_0$ width, due to the laser period is on a time scale significantly shorter than the SRS growth time. For the normal laser case, the intensity keeps stable over time, with the reflectivity rises at $t=700\tau_0$. In the subsequent non-linear stage, the reflectivity fluctuates with time. The corresponding electron plasma waves are illustrated in Figure 2\;(c), where many EPW packets are generated near the left boundary and convect forward. Electrons are trapped in these periodical wave packets, which leads to local pump depletion, and further leads to the fluctuation of reflectivity. This mechanism has been discussed in detail by \emph{Winjum} \etal\cite{RN23}. For the broadband laser case, the reflectivity rises delayed to $t=1500\tau_0$, which is consistent with the corresponding EPW shows in Figure 2\;(d). The delay is in line with the expectation of linear theory and has been observed in several previous studies by \emph{Zhao} \etal and \emph{Zhou} \etal\cite{RN13,RN18}. \par

The previous works use large bandwidth laser (over $5\%$) and focus on the frequency effect of broadband laser\cite{RN19}. However, based on the performance of current laser facilities, we are more interested in the evolution of parametric instabilities driven by the laser with relatively narrow bandwidth. We found that the reflectivity driven by a broadband laser has an entirely different growth pattern compared with the normal laser case. A series of steep reflectivity spikes are observed after saturation in Figure 2\;(b), shows an intermittent burst pattern. Note that the intensity of broadband laser significantly changes over time, as we have discussed in Section 2. This variation is a key factor in the SRS growth process. 
For a homogeneous plasma, the linear coupled mode equations are:

\begin{equation}
   \left(\frac{\partial^2}{\partial t^2}-c^2\frac{\partial^2}{\partial x^2}+\omega_{pe}^2 \right)a_s=-\omega_{pe}^2 a_p a_0, \label{3}
\end{equation}

\begin{equation}
   \left(\frac{\partial^2}{\partial t^2}-3 v_{the}^2\frac{\partial^2}{\partial x^2}+\omega_{pe}^2 \right)a_p=c^2\frac{\partial^2}{\partial x^2}(a_s a_0), \label{4}
\end{equation}

and the growth rate is:

\begin{equation}
   \gamma_0=\frac{kv_{os}}{4}\sqrt{\frac{\omega^2_{pe}}{\omega_{ek}(\omega_0-\omega_{ek})}},\label{5}
\end{equation}

where $k(\omega_{ek})$ are the frequency (wave number) of electron plasma wave, and $v_{os}$ is the oscillatory velocity of an electron in the laser. Since the growth rate of SRS is in proportion to the laser intensity, one can find a relation between the high-intensity pulses and reflectivity spikes, which indicate the intensity variation dominates the SRS intermittent burst in the broadband laser case. It should be mentioned that there is a time interval between the peaks of reflectivity and laser intensity due to the propagation of incident and scatter waves in plasma.

\subsection{Transition from Convective to Absolute Instability}
It is well known that the SRS growth behavior in an absolute instability is significantly different from that in a convective instability. In this section, we will show that the transition from convective to absolute instability is likely to occur due to the synergism between the high-intensity pulses in the broadband laser case. Figure 2\;(b) shows that the reflectivity has a strong burst and reaches a peak value $r_{max}$ at $t=2800\tau_0$, which is several times higher than that of the subsequent spikes. By contrast, the corresponding pulse intensity is slightly different from other pulses. In order to explain this non-linear behavior, the transition from convective to absolute instability has to be taken into account.\par

When the reflectivity rises sharply at $2800\tau_0$, a bright band is observed in Figure 2\;(d), marked with a white arrow. The bright band structure implies the SRS rises sharply in different positions of space at a certain time, which is consistent with the signs of absolute instability. More evidence of the transition is shown in Figure 3, where a high-intensity pulse is incidents into the plasma at $t=2750\tau_0$, followed by the energy of EPWs increases sharply. Before that, the energy of EPWs for both the normal laser case and the broadband laser case has been saturated at the same level. These signs indicate that one high-intensity pulse may exceed the absolute instability threshold and lead to an SRS burst.\par

\begin{figure}
   \centering
   \includegraphics[width=0.96\textwidth]{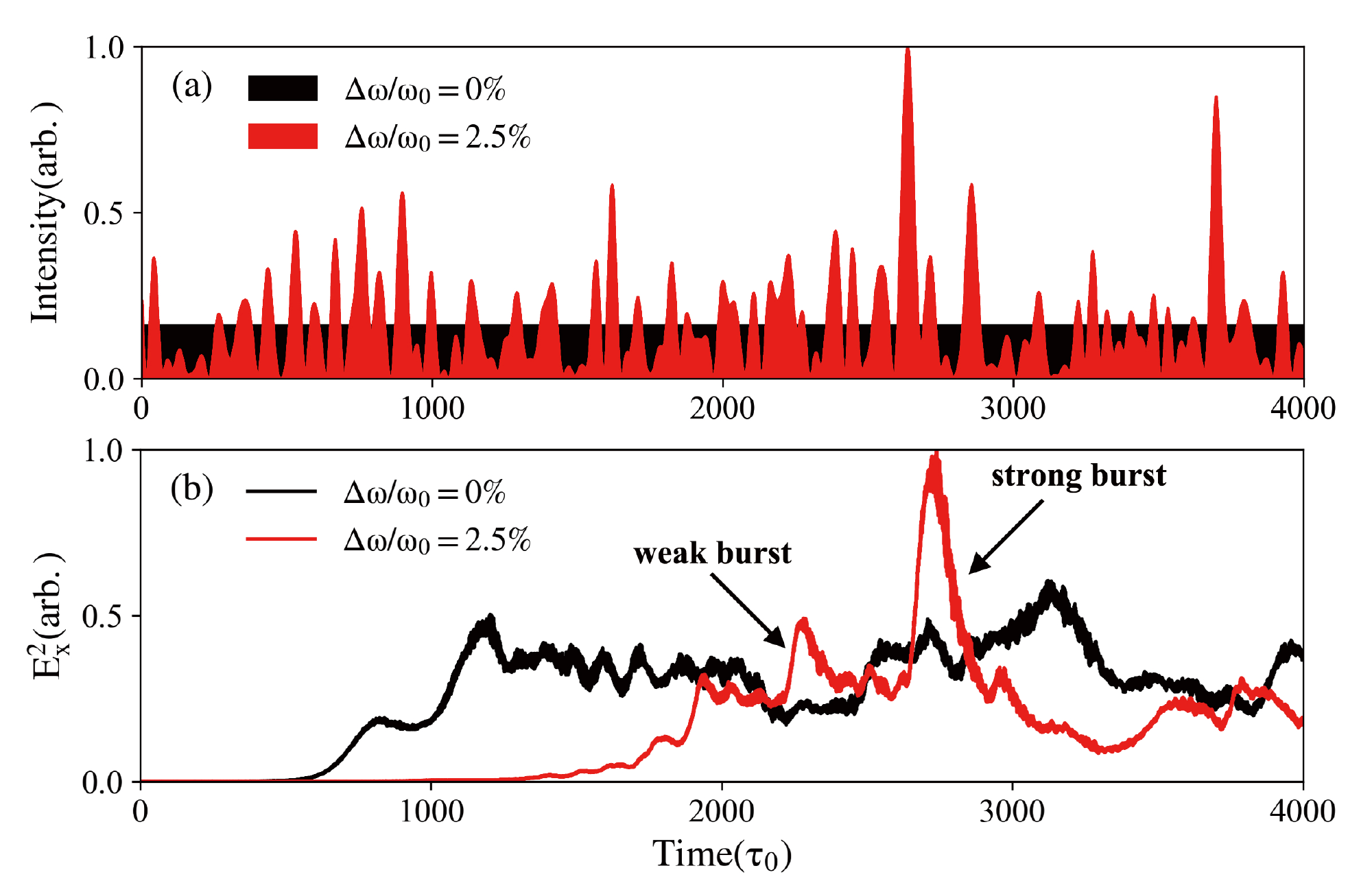}
   \caption{Temporal envelopes of the incident laser intensity for the normal laser case and the broadband laser case (a), normalized to the $I_{max}$. The energy of electron plasma waves ($E_x^2$) vs. time, normalized to the max energy $E_{max}^2$ in the broadband laser case. }
   \label{fig:3}
\end{figure}

For a homogeneous infinite plasma, the absolute instability threshold of SRS is\cite{RN24}:

\begin{equation}
   \frac{\gamma^2_{abs}}{v_l v_s}=(\frac{\pi}{L})^2+\frac{1}{4}(\frac{\nu_l}{v_l})^2+\frac{1}{4}(\frac{\nu_s}{v_s})^2.\label{6}
\end{equation}

Here, $\gamma_{abs}$ is the absolute instability threshold of SRS, $v_l$ and $v_s$ are the group velocity of the EPW and the scattering light, $\nu_l$ and $\nu_s$ are the damping rates of the EPW and the scattering light. $L$ is the length of the simulation box. Considering the parameters in our simulations, the first and third terms on the right of equation (6) can be ignored since the simulation box is long ($100\lambda_0$) and the initial electron temperature is high ($3keV$). Thus equation (6) can be simplified as:

\begin{equation}
   \frac{\gamma^2_{abs}}{v_l v_s}=\frac{1}{4}(\frac{\nu_l}{v_l})^2.\label{7}
\end{equation}

\par For $k\lambda_d=0.32$ in our simulations, the EPW’s phase velocity is near the electron thermal velocity, thus the Landau damping ($\nu_l$) is high. The threshold for the absolute instability calculated by equation (7) is $I_{abs}=4.3\times10^{16} W/cm^2$,  which is still too high for the pulse at $2750\tau_0$ with a maximum intensity of $1.6\times10^{16} W/cm^{2}$. Furthermore, a pulse with similar intensity is incident into the plasma at a later time $t = 3800\tau_0$, whereas the SRS has only a weak response shown in Figure 3\;(b). Therefore, an important question arises: What caused the strong burst of SRS?
\par
Our analysis below provides a definitive answer to this question. The physics of the strong burst is straightforward: When a train of pulses incident into the simulation box, the former pulses will enhance the SRS excited by the latter pulse through the following mechanism. (1) The former pulses incident into plasma and generate EPWs in the plasma slab's shallow layer. (2) These EPWs propagate in the laser direction with a velocity $v_{EPW}$. Many electrons are trapped in these EPWs and co-propagate with them. The electron trapping can flatten the electron distribution function around the phase velocity of EPWs, which causes the reduction of Landau damping so that the threshold of absolute instability decreases.\cite{RN24,YIN1} (3) A latter high-intensity pulse incidents into plasma. When the pulse propagates to the region where Landau damping decreases, its intensity exceeds the absolute instability threshold. Thus, the transition from convective to absolute instability is triggered, leading to a burst of instantaneous reflectivity and producing many hot electrons. (4) After the high-intensity pulse leaves the plasma, the instability decreases rapidly.

\begin{figure}
   \centering
   \includegraphics[width=0.96\textwidth]{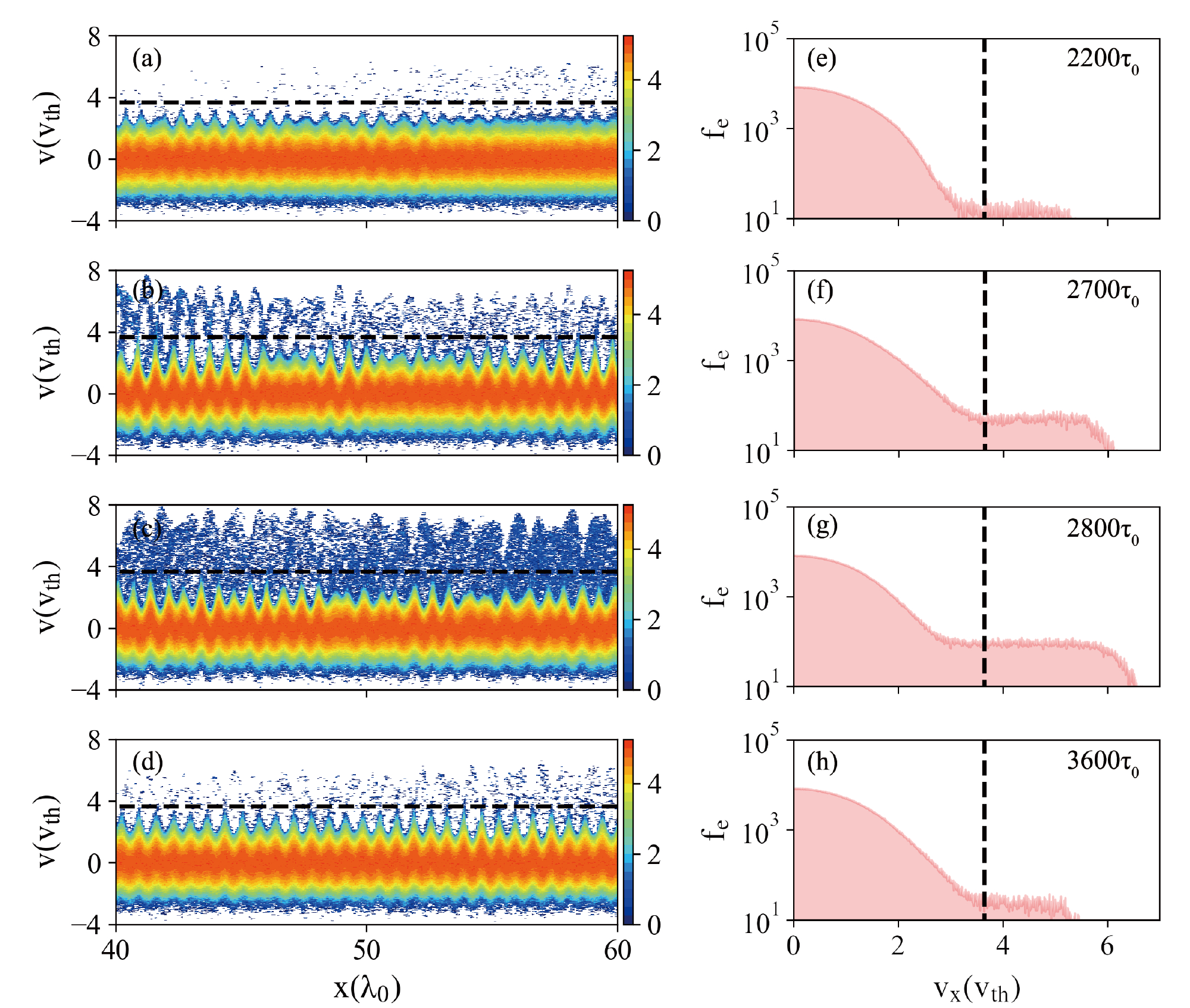}
   \caption{Electron momentum phase space snapshots in the central region ($40\lambda_0-60\lambda_0$) for the broadband laser case at $t=2200\tau_0$ (a), $t=2700\tau_0$ (b), $t=2800\tau_0$ (c) and $t=3600\tau_0$ (d). Electron velocity distribution function snapshots at the corresponding time (e-h). The dash lines represent the velocity of EPWs. }
   \label{fig:4}
\end{figure}

In our simulations, a relatively weak SRS burst was excited at $t=2300\tau_0$, as shown in Figure 2\;(b) and Figure 3\;(b). This burst generated EPWs in the shallow layer of the plasma ($10\lambda_0-30\lambda_0$), as shown in Figure 2\;(d). A white arrow is used to mark the trajectories of these EPWs, and the slope of the arrow represents their group velocity $v_{EPW}$. Before the strong burst at $2800\tau_0$, these EPWs propagated to the center of the plasma slab ($40\lambda_0-60\lambda_0$) with the trapped electrons. Figure 4 shows that the trapped electrons significantly change the electron momentum phase space and distribution function in the central region of the plasma slab. When the weak burst had not been excited, only a small number of electrons was trapped at $2200\tau_0$. After $400\tau_0$, the EPWs and hot electrons produced by the weak burst reached the central region and flattened the electron distribution function, significantly reducing the Landau damping. Thus, the absolute threshold drops from $I_{abs}=4.3\times10^{16} W/cm^2$ to $I_{abs}=4.8\times10^{15} W/cm^2$. Therefore, the latter high-intensity pulse triggered the transition to absolute instability, which excited a strong burst of instantaneous reflectivity and produced many hot electrons at $2800\tau_0$, as shown in Figure 2\;(b), Figure 4\;(c) and (g). Although the generation of hot electrons is particularly problematic for indirect ICF, these hot electrons can enhance shock drive performance by enhanced ablation pressures for shock ignition\cite{SI1}. As a comparison, almost no high-intensity pulses incident the plasma before the pulse at $3800\tau_0$, and there are not enough electrons be trapped in the central region. Thus, the threshold does not fall below the pulse intensity, and the SRS has only a weak response. In a word, the synergism between pulses is the key factor leading to SRS burst.

\subsection{Effect of Laser Bandwidth}

The theoretical results obtained by \emph{Thomson} \etal have shown that if the laser power is spread over a wide frequency region $\Delta\omega\gg \gamma$, the growth rates would decrease significantly\cite{RN11}. These results were verified by \emph{Zhao} \etal\cite{RN13} and \emph{Follet} \etal\cite{RN16}. However, a new understanding of this phenomenon could be proposed by considering the average pulse duration of the broadband laser. We define the characteristic growth time of BSRS as:

\begin{equation}
   \tau_{BSRS}=1/\gamma_{max},\label{8}
\end{equation}

where the $\gamma_{max}$ is the maximum growth rate of BSRS. By comparing equation (8) with equation (2) and considering $\Delta\omega=2\pi\Delta\nu$, one can get $\tau_c=1/\Delta\nu$. When the bandwidth of the laser is larger than the maximum growth rate of SRS, the average duration of the pulses will be less than the characteristic growth time of SRS. This result means that the growth process of SRS will be interrupted frequently, so that SRS will be suppressed. To confirm this inference, we performed simulations using broadband lasers with different bandwidths. We use three bandwidths, $2.5\%$, $5.0\%$ and $10.0\%$, corresponding to the average duration $46.8fs$, $23.4fs$ and $11.7fs$. Note that the characteristic time in our simulation is $44.3fs$. The simulation results are shown in Figure 5. In the normal laser case, the energy of scattered light increases sharply at an early time and saturates at a high level. In the $\Delta\omega/\omega_0=2.5\%$ case, the scattered light increases later, through the increasing process is tortuous due to the intermittent burst, and the energy reaches a similar level with normal laser case eventually. When the bandwidth $\Delta\omega/\omega_0> 5\%$, the energy of scattered light decays to a thousandth of the previous simulations. These results have shown that SRS is suppressed in the large bandwidth simulations. The suppression effect is due to the low average duration, which frequently interrupts the SRS growth process. Meanwhile, large bandwidth leads to the decrease of effective laser energy, thus decreasing the growth rate of SRS.

\begin{figure}
    \centering
    \includegraphics[width=0.96\textwidth]{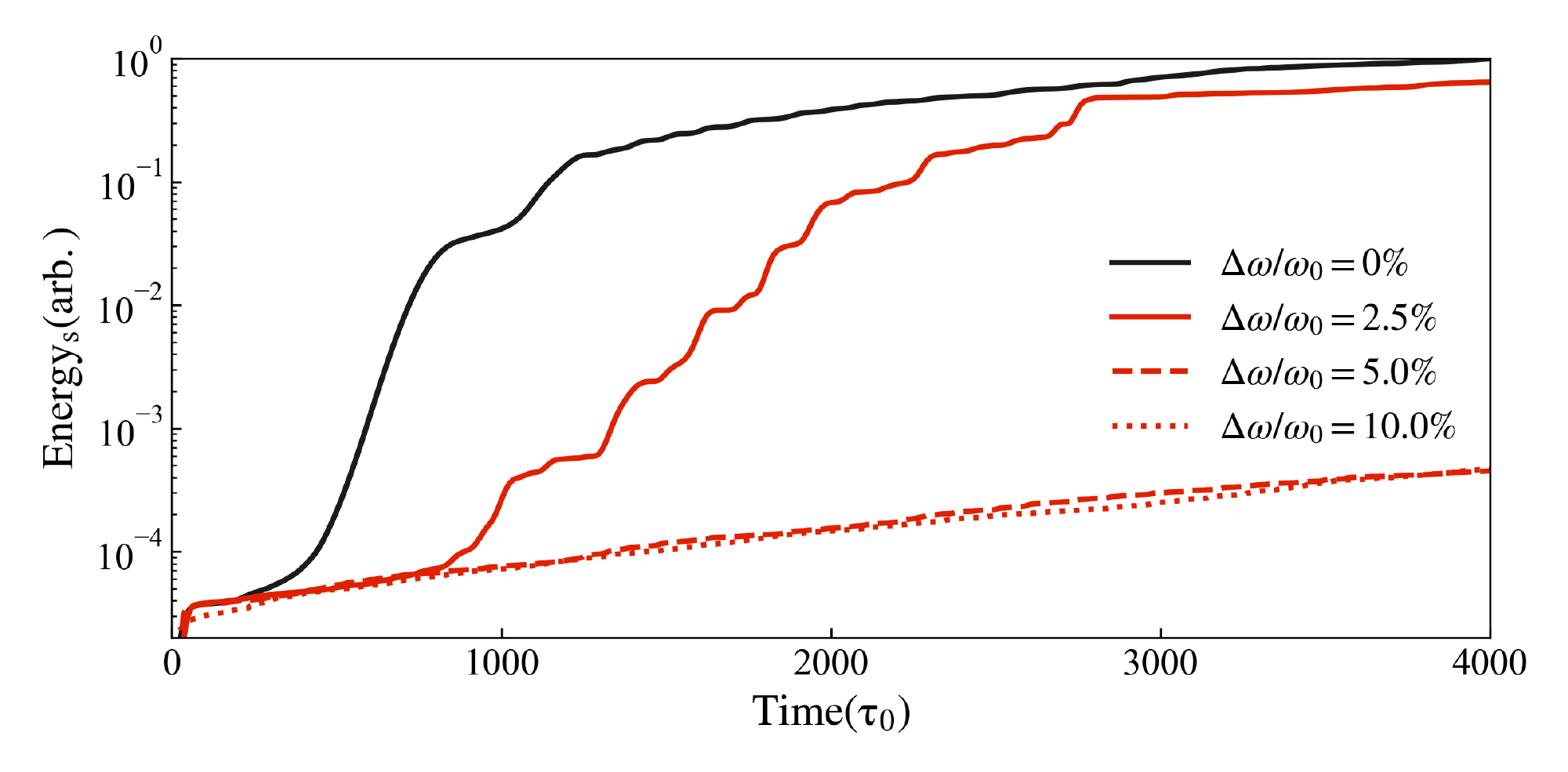}
    \caption{Energy of scattered light vs time. Note that logarithmic coordinates are used. }
    \label{fig:5}
 \end{figure}

\subsection{Intermittent Excitation of TPD}

Understanding and controlling the TPD growth process is a critical part of ICF schemes. Different from SRS, TPD always occurs near $0.25n_c$ and can only be investigated by 2D or 3D simulations. However, the TPD growth rate is also in proportion to the laser intensity, which means the TPD driven by a broadband laser may also burst intermittently with the same mechanism as discussed above. A set of 2D PIC simulations have been performed to study the evolution of TPD driven by a broadband laser. One simulation uses a broadband laser with a bandwidth $\Delta\omega/\omega_0=2.5\%$, and the other uses a normal laser as a control. Both lasers are linearly polarized with the wavelength $\lambda_0=351nm$ and the intensity $I_0=4.0\times 10^{15} W/cm^2$. The simulation box contains an un-dense plasma slab with sharp edges normal to the direction of the laser. The slab length is $72\lambda_0$, and on each side of the plasma slab, there is a $4\lambda_0$ vacuum region. The box width is $30\lambda_0$ in the transverse direction. The plasma has a linear density profile from $0.17n_c$ to $0.33n_c$, and an electron temperature $T_e=3keV$. There are 20 grid points per $\lambda_0$ and 100 particles per grid inside the slab, which yield a total of $9.6\times10^8$ particles. One simulation lasts $4000\tau_0$, or equivalently $4.7ps$.\par

\begin{figure}
   \includegraphics[width=0.96\textwidth]{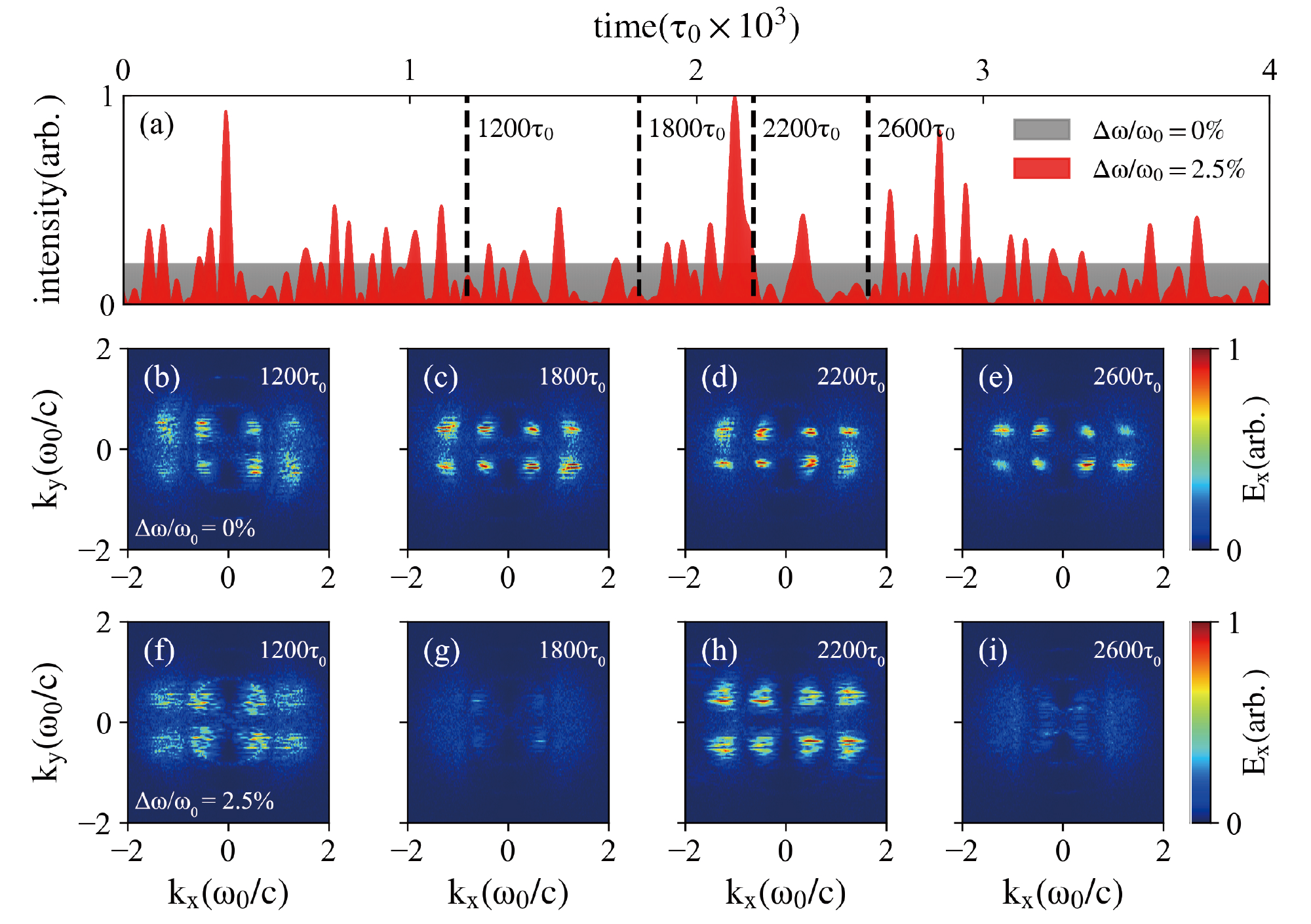}
   \caption{The intensity envelops for a normal laser and a broadband laser (a) in arbitrary units. Spectra of $E_x$ in the $k_y-k_x$ spaces for a normal laser (b-e) and a broadband laser (f-i) at four different times. }
   \label{fig:6}
\end{figure}

Simulation results are illustrated in Figure 6. As can be seen from Figure 6\;(a), the broadband laser has a maximum intensity at $t=2100\tau_0$, and the dash lines represent four different times. Figure 6\;(b-e) show the spectra of $E_x$ at the corresponding time for the normal laser case and Figure 6\;(f-i) for the broadband laser case. Typical $E_x (k)$ structures of TPD have been observed in both simulations at $t=1200\tau_0$. However, after saturation, the TPD growth process in a broadband laser case is quite different from the normal laser case, which is like SRS mentioned in Section 3.1. For the normal laser case, the intensity of EPWs produced by TPD keeps stable over time with constant laser intensity. For the broadband laser case, the intensity of EPWs changes greatly with a significant variation shown in Figure 6\;(f-i). The pulses intensity remains at a low level during $1200\tau_0$ to $1800\tau_0$, leading to a rapid decrease of TPD. A	high-intensity pulse incident into the simulation box at $t=2100\tau_0$, which cause a sharply rebound of TPD at $t=2200\tau_0$. In the next $400\tau_0$, the decrease of pulse intensity leads to a re-decrease of TPD. The above discussions show that the variation of laser intensity also dominates the growth process of TPD. These random high-intensity pulses lead to intermittent bursts of instabilities in both SRS and TPD simulations.

\section{Summary}
In summary, our work have shown a new evolution pattern for both SRS and TPD driven by a broadband laser. The change of coherence of different frequency beamlets leads to the intensity variation of the broadband laser. Consequently, it causes an intermittent excitation of parametric instabilities, observed in both 1D and 2D simulations. The intensity variation is treated as a series of pulses with an average duration $\tau_{avg}=2\pi/\Delta\omega$, which varies inversely with the bandwidth of the broadband laser. With increasing the laser bandwidth, the pulse duration $\tau_{avg}$ falls below the characteristic time of SRS $\tau_{BSRS}$, and the reflectivity can be suppressed at a low level. However, the synergism of these pulses was also observed with the relatively narrow bandwidth. A former pulse can generate EPWs with trapped electrons and decreases the threshold of absolute instability. Thus a later high-intensity pulse can trigger the non-linear transition from convective to absolute SRS and enhance the intermittent excitation. Important topics, such as quantitative analysis of the synergistic effects of these high-intensity pulses with statistical methods, should be explored in future works.

\ack
This work was supported by the National Natural Science Foundation of China (Grants No. 11975055, 12075033, 11905015 and No. U1730449 (NSAF)),and the National Key Programme for $S\&T$ Research and Development (Grant No. 2016YFA0401100). The simulations were performed on the Tianhe-2 supercomputer (China). I am grateful to Y. Zhao, H. Y. Zhou and C. Y. Zheng for their valuable suggestions. Also I wish to thank my colleagues for their many courtesies.

\section*{References}
\bibliographystyle{unsrt}
\bibliography{rk}

\begin{thebibliography}{10}

\bibitem{RN0}
R.~Betti and O.~A. Hurricane.
\newblock Inertial-confinement fusion with lasers.
\newblock {\em Nature Physics}, 12(5):435--448, 2016.

\bibitem{RN1}
Mark Buchanan.
\newblock Evolution of fusion.
\newblock {\em Nature Physics}, 15(7):620--620, 2019.

\bibitem{RN2}
Tao Gong, Liang Hao, Zhichao Li, Dong Yang, Sanwei Li, Xin Li, Liang Guo,
  Shiyang Zou, Yaoyuan Liu, Xiaohua Jiang, Xiaoshi Peng, Tao Xu, Xiangming Liu,
  Yulong Li, Chunyang Zheng, Hongbo Cai, Zhanjun Liu, Jian Zheng, Zhebin Wang,
  Qi~Li, Ping Li, Rui Zhang, Ying Zhang, Fang Wang, Deen Wang, Feng Wang,
  Shenye Liu, Jiamin Yang, Shaoen Jiang, Baohan Zhang, and Yongkun Ding.
\newblock Recent research progress of laser plasma interactions in shenguang
  laser facilities.
\newblock {\em Matter and Radiation at Extremes}, 4(5):055202, 2019.

\bibitem{NF1}
J.~L. Kline, S.~H. Batha, L.~R. Benedetti, D.~Bennett, S.~Bhandarkar,
  L.~F.~Berzak Hopkins, J.~Biener, M.~M. Biener, R.~Bionta, E.~Bond,
  D.~Bradley, T.~Braun, D.~A. Callahan, J.~Caggiano, C.~Cerjan, B.~Cagadas,
  D.~Clark, C.~Castro, E.~L. Dewald, T.~Döppner, L.~Divol, R.~Dylla-Spears,
  M.~Eckart, D.~Edgell, M.~Farrell, J.~Field, D.~N. Fittinghoff,
  M.~Gatu~Johnson, G.~Grim, S.~Haan, B.~M. Haines, A.~V. Hamza, E.~P. Hartouni,
  R.~Hatarik, K.~Henderson, H.~W. Herrmann, D.~Hinkel, D.~Ho, M.~Hohenberger,
  D.~Hoover, H.~Huang, M.~L. Hoppe, O.~A. Hurricane, N.~Izumi, S.~Johnson,
  O.~S. Jones, S.~Khan, B.~J. Kozioziemski, C.~Kong, J.~Kroll, G.~A. Kyrala,
  S.~LePape, T.~Ma, A.~J. Mackinnon, A.~G. MacPhee, S.~MacLaren, L.~Masse,
  J.~McNaney, N.~B. Meezan, J.~F. Merrill, J.~L. Milovich, J.~Moody, A.~Nikroo,
  A.~Pak, P.~Patel, L.~Peterson, E.~Piceno, L.~Pickworth, J.~E. Ralph, N.~Rice,
  H.~F. Robey, J.~S. Ross, J.~R. Rygg, M.~R. Sacks, J.~Salmonson, D.~Sayre,
  J.~D. Sater, M.~Schneider, M.~Schoff, S.~Sepke, R.~Seugling, V.~Smalyuk,
  B.~Spears, M.~Stadermann, W.~Stoeffl, D.~J. Strozzi, R.~Tipton, C.~Thomas,
  P.~L. Volegov, C.~Walters, M.~Wang, C.~Wilde, E.~Woerner, C.~Yeamans, S.~A.
  Yi, B.~Yoxall, A.~B. Zylstra, J.~Kilkenny, O.~L. Landen, W.~Hsing, et~al.
\newblock Progress of indirect drive inertial confinement fusion in the united
  states.
\newblock {\em Nuclear Fusion}, 59(11):112018, 2019.

\bibitem{RN3}
Chuan~Sheng Liu and Vijai~Kumar Tripathi.
\newblock {\em Interaction of electromagnetic waves with electron beams and
  plasmas}.
\newblock World Scientific, 1994.

\bibitem{RN4}
William~L. Kruer.
\newblock {\em The Physics of Laser Plasma Interactions}.
\newblock crc Press, 2019.

\bibitem{NF2}
K.~Q. Pan, S.~E. Jiang, Q.~Wang, L.~Guo, S.~W. Li, Z.~C. Li, D.~Yang, C.~Y.
  Zheng, B.~H. Zhang, and X.~T. He.
\newblock Two-plasmon decay instability of the backscattered light of
  stimulated raman scattering.
\newblock {\em Nuclear Fusion}, 58(9):096035, 2018.

\bibitem{NF3}
C.~Z. Xiao, H.~B. Zhuo, Y.~Yin, Z.~J. Liu, C.~Y. Zheng, and X.~T. He.
\newblock Transition from two-plasmon decay to stimulated raman scattering
  under ignition conditions.
\newblock {\em Nuclear Fusion}, 60(1):016022, 2019.

\bibitem{RN5}
R.~K. Kirkwood, J.~D. Moody, J.~Kline, E.~Dewald, S.~Glenzer, L.~Divol,
  P.~Michel, D.~Hinkel, R.~Berger, E.~Williams, J.~Milovich, L.~Yin, H.~Rose,
  B.~MacGowan, O.~Landen, M.~Rosen, and J.~Lindl.
\newblock A review of laser–plasma interaction physics of indirect-drive
  fusion.
\newblock {\em Plasma Physics and Controlled Fusion}, 55(10):103001, 2013.

\bibitem{RN6}
R.~Yan, C.~Ren, J.~Li, A.~V. Maximov, W.~B. Mori, Z.~M. Sheng, and F.~S. Tsung.
\newblock Generating energetic electrons through staged acceleration in the
  two-plasmon-decay instability in inertial confinement fusion.
\newblock {\em Phys Rev Lett}, 108(17):175002, 2012.

\bibitem{RN7}
B.~J. Winjum, F.~S. Tsung, and W.~B. Mori.
\newblock Mitigation of stimulated raman scattering in the kinetic regime by
  external magnetic fields.
\newblock {\em Physical Review E}, 98(4):043208, 2018.

\bibitem{RN8}
R.~H. Lehmberg and S.~P. Obenschain.
\newblock Use of induced spatial incoherence for uniform illumination of laser
  fusion targets.
\newblock {\em Optics Communications}, 46(1):27--31, 1983.

\bibitem{RN9}
J.~D. Moody, B.~J. MacGowan, J.~E. Rothenberg, R.~L. Berger, L.~Divol, S.~H.
  Glenzer, R.~K. Kirkwood, E.~A. Williams, and P.~E. Young.
\newblock Backscatter reduction using combined spatial, temporal, and
  polarization beam smoothing in a long-scale-length laser plasma.
\newblock {\em Phys Rev Lett}, 86(13):2810--3, 2001.

\bibitem{RN10}
Ping Li, Sai Jin, Runchang Zhao, Wei Wang, Fuquan Li, Mingzhong Li, Jingqin Su,
  and Xiaofeng Wei.
\newblock The special shaped laser spot for driving indirect-drive hohlraum
  with multi-beam incidence.
\newblock {\em High Power Laser Science and Engineering}, 5, 2017.

\bibitem{RN11}
J.~J. Thomson.
\newblock Effects of finite-bandwidth driver on the parametric instability.
\newblock {\em Physics of Fluids}, 17(8):1608--1613, 1974.

\bibitem{RN12}
D.~Pesme, R.~L. Berger, E.~A. Williams, A.~Bourdier, and A.~Bortuzzo-Lesne.
\newblock A statistical description of parametric instabilities with an
  incoherent pump, 2007.

\bibitem{RN13}
Jun Zheng, LuLe Yu, Yao Zhao, Min Chen, SuMing Weng, and ZhengMing Sheng.
\newblock Effects of large laser bandwidth on stimulated raman scattering
  instability in underdense plasma.
\newblock {\em SCIENTIA SINICA Physica, Mechanica \& Astronomica},
  45(3):035201--035201, 2015.

\bibitem{RN14}
Yao Zhao, Suming Weng, Min Chen, Jun Zheng, Hongbin Zhuo, Chuang Ren, Zhengming
  Sheng, and Jie Zhang.
\newblock Effective suppression of parametric instabilities with decoupled
  broadband lasers in plasma.
\newblock {\em Physics of Plasmas}, 24(11):112102, 2017.

\bibitem{RN15}
Yao Zhao, Suming Weng, Zhengming Sheng, and Jianqiang Zhu.
\newblock Suppression of parametric instabilities in inhomogeneous plasma with
  multi-frequency light.
\newblock {\em Plasma Physics and Controlled Fusion}, 61(11):115008, 2019.

\bibitem{RN16}
R.~K. Follett, J.~G. Shaw, J.~F. Myatt, C.~Dorrer, D.~H. Froula, and J.~P.
  Palastro.
\newblock Thresholds of absolute instabilities driven by a broadband laser.
\newblock {\em Physics of Plasmas}, 26(6):062111, 2019.

\bibitem{RN17}
R.~K. Follett, J.~G. Shaw, J.~F. Myatt, H.~Wen, D.~H. Froula, and J.~P.
  Palastro.
\newblock Thresholds of absolute two-plasmon-decay and stimulated raman
  scattering instabilities driven by multiple broadband lasers.
\newblock {\em Physics of Plasmas}, 28(3):032103, 2021.

\bibitem{RN18}
H.~Y. Zhou, C.~Z. Xiao, D.~B. Zou, X.~Z. Li, Y.~Yin, F.~Q. Shao, and H.~B.
  Zhuo.
\newblock Numerical study of bandwidth effect on stimulated raman
  backscattering in nonlinear regime.
\newblock {\em Physics of Plasmas}, 25(6):062703, 2018.

\bibitem{GAO1}
Yong Cui, Yanqi Gao, Daxing Rao, Dong Liu, Fujian Li, Lailin Ji, Haitao Shi,
  Jiani Liu, Xiaohui Zhao, Wei Feng, Lan Xia, Jia Liu, Xiaoli Li, Tao Wang,
  Weixin Ma, and Zhan Sui.
\newblock High-energy low-temporal-coherence instantaneous broadband pulse
  system.
\newblock {\em Opt. Lett.}, 44(11):2859--2862, Jun 2019.

\bibitem{RN20}
Yanqi Gao, Yong Cui, Lailin Ji, Daxing Rao, Xiaohui Zhao, Fujian Li, Dong Liu,
  Wei Feng, Lan Xia, Jiani Liu, Haitao Shi, Pengyuan Du, Jia Liu, Xiaoli Li,
  Tao Wang, Tianxiong Zhang, Chong Shan, Yilin Hua, Weixin Ma, Xun Sun,
  Xianfeng Chen, Xiuguang Huang, Jian Zhu, Wenbing Pei, Zhan Sui, and Sizu Fu.
\newblock Development of low-coherence high-power laser drivers for inertial
  confinement fusion.
\newblock {\em Matter and Radiation at Extremes}, 5(6):065201, 2020.

\bibitem{RN19}
C.~Benedetti, C.~B. Schroeder, E.~Esarey, and W.~P. Leemans.
\newblock Plasma wakefields driven by an incoherent combination of laser
  pulses: A path towards high-average power laser-plasma accelerators.
\newblock {\em Physics of Plasmas}, 21(5):056706, 2014.

\bibitem{RN21}
Joseph~W Goodman.
\newblock {\em Statistical optics}.
\newblock John Wiley \& Sons, 2015.

\bibitem{RN22}
Hong-bo Cai, Xin-xin Yan, Pei-lin Yao, and Shao-ping Zhu.
\newblock Hybrid fluid–particle modeling of shock-driven hydrodynamic
  instabilities in a plasma.
\newblock {\em Matter and Radiation at Extremes}, 6(3):35901, 2021.

\bibitem{RN23}
B.~J. Winjum, J.~E. Fahlen, F.~S. Tsung, and W.~B. Mori.
\newblock Effects of plasma wave packets and local pump depletion in stimulated
  raman scattering.
\newblock {\em Phys Rev E Stat Nonlin Soft Matter Phys}, 81(4 Pt 2):045401,
  2010.

\bibitem{RN24}
Y.~X. Wang, Q.~Wang, C.~Y. Zheng, Z.~J. Liu, C.~S. Liu, and X.~T. He.
\newblock Nonlinear transition from convective to absolute raman instability
  with trapped electrons and inflationary growth of reflectivity.
\newblock {\em Physics of Plasmas}, 25(10):100702, 2018.

\bibitem{YIN1}
L.~Yin, B.~J. Albright, H.~A. Rose, K.~J. Bowers, B.~Bergen, and R.~K.
  Kirkwood.
\newblock Self-organized bursts of coherent stimulated raman scattering and hot
  electron transport in speckled laser plasma media.
\newblock {\em Phys. Rev. Lett.}, 108:245004, Jun 2012.

\bibitem{SI1}
L.~J. Perkins, R.~Betti, K.~N. LaFortune, and W.~H. Williams.
\newblock Shock ignition: A new approach to high gain inertial confinement
  fusion on the national ignition facility.
\newblock {\em Phys. Rev. Lett.}, 103:045004, Jul 2009.

\end{thebibliography}

\end{document}